\documentclass[prl,twocolumn,amsmath,amssymb,groupedaddress]{revtex4}
\usepackage{dcolumn}% Align table columns on decimal point
\usepackage{graphicx,subfigure}% Include figure files
\usepackage{bm}% bold math
\usepackage{verbatim}
\usepackage{amsmath}
\usepackage{amssymb}
\usepackage[T1]{fontenc}
\usepackage{ae,aecompl}
\usepackage{appendix}
\usepackage{float}
\usepackage{color}

\newcommand{\pt}{{\partial}}

\newcommand{\rv}{{\bf r}}

\newcommand{\thv}{{\bm{\theta}}}
\newcommand{\chiv}{{\bm{\chi}}}
\newcommand{\av}{{\bf a}}

\newcommand{\A}{{\bf A}}

\newcommand{\J}{{\bf J}}

\newcommand{\E}{{\bf E}}

\newcommand{\xv}{{\bf x}}

\newcommand{\ev}{{\bf e}}

\newcommand{\jv}{{\bf j}}

\newcommand{\oh}{{\frac{1}{2}}}

\newcommand{\cH}{{\mathcal H}}
\newcommand{\cL}{{\mathcal L}}
\newcommand{\grad}{{\bm{\nabla}}}

\newcommand{\be}{\begin{equation}}
\newcommand{\ee}{\end{equation}}
\newcommand{\bea}{\begin{eqnarray}}
\newcommand{\eea}{\end{eqnarray}}
\newcommand{\bse}{\begin{subequations}}
\newcommand{\ese}{\end{subequations}}
\def\rf#1{(\ref{#1})}

\begin{document}
\title{Lifshitz gauge duality}
\author{Leo Radzihovsky}
\affiliation{
Department of Physics and Center for Theory of Quantum Matter\\
University of Colorado, Boulder, CO 80309}
\date{\today}
\email{radzihov@colorado.edu}
%\date{\today}

\begin{abstract}
  Motivated by a variety of realizations of the compact Lifshitz model
  I derive its fractonic gauge dual. The resulting U(1) vector gauge
  theory efficiently and robustly encodes the restricted mobility of
  its dipole conserving charged matter and the corresponding
  topological vortex defects.  The gauge theory provides a transparent
  formulation of the three phases of the Lifshitz model and gives a
  field theoretic formulation of the associated two-stage Higgs
  transitions.
\end{abstract}

\maketitle

%\noindent{\em Introduction.}

{\em Introduction and motivation.} Recently, there have been much
interest in systems with fine-tuned generalized global symmetries and
their fractonic gauge duals.\cite{LRGromovRMP} One of the simplest is
the Lifshitz model (and its $m$-Lifshitz
generalization\cite{mLifshitz}), that describes a diverse number of
physical systems.  Its classical realizations date back a half century
in studies of Goldstone modes of cholesteric, smectic, and columnar
liquid crystals, tensionless membranes and nematic
elastomers\cite{LRcholesteric, JWSmembranes, RTtubulePRL, RTtubulePRE,
  LRmembranes, LRbuckling, MoessnerMembrane,XRelastomers}, and many
other soft-matter phases exhibiting rich phenomenology
\cite{ProstDeGennes,ChaikinLubensky}.
%that is at the heart of the modern
%liquid crystal display technology.
%Other classical
%examples include dynamics of tensionless
%membranes\cite{JerusalemWS,MoessnerMembrane}.

Quantum realizations of the Lifshitz model include Hall striped states
of a two-dimensional electron gas at half-filled high Landau
levels\cite{EisensteinQSm,Fogler,Moessner,FradkinKivelsonQHsm,FisherMacdonald,LR_Dorsey},
striped spin and charge states of weakly doped correlated quantum
magnets\cite{TranquadaStripes,KivelsonStripes}, critical theory of the
RVB - VBS transition\cite{MoessnerSondhiFrandkin2000,
  VishwanathBalentsSenthil2004,FradkinHuse2004}, ferromagnetic
transition in one-dimensional spin-orbit-coupled
metals\cite{KoziiPRB17}, the putative Fulde-Ferrell-Larkin-Ovchinnikov
(FFLO) paired superfluid\cite{FF, LO} in imbalanced degenerate atomic
gases\cite{LR_VishwanathPRL,LRpra}, and spin-orbit coupled Bose
condensates.\cite{LR_ChoiPRL,HuiZhai}, as well helical states of
bosons or spins on a frustrated lattice\cite{helicalBosonsSMR22}.

The most notable feature of the 3d classical and 2+1d quantum Lifshitz
model is its enlarged ``tilt'' or dipolar symmetry and the concomitant
logarithmic ``roughness'', $\phi_{rms}^2\sim \log L$ of its Goldstone
mode $\phi$ (akin to the XY model in two dimensions), that leads to
its power-law correlated, quasi-long-range ordered state for the
matter field $e^{i \phi}$. Depending on the nature of its physical
realization, the enlarged symmetry may result from fine-tuning to a
critical point -- as e.g., in RVB - VBS
\cite{MoessnerSondhiFrandkin2000,
  VishwanathBalentsSenthil2004,FradkinHuse2004},
paramagnetic-ferromagnetic in spin-orbit-coupled
metals\cite{KoziiPRB17} and a membrane buckling\cite{LRbuckling} phase
transitions, or is dictated by an underlying symmetry -- as e.g.,
``target-space'' rotational invariance of smectic, columnar,
cholesteric, and tensionless membrane ordered
phases.\cite{Grinstein86, LRpra, LRcholesteric, JWSmembranes,
  RTtubulePRL, RTtubulePRE, LRmembranes} In these realizations the
nonlinearities (elastic in the context of smectics and membrane
states) become relevant for $d < d_c$ ($d_c = 3$ and $d_c=5/2$ for the
classical smectic\cite{Grinstein86, LRpra} and columnar
states\cite{columnarVglassPRL,columnarBGprl}, respectively), leading
to universal ``critical phases''.

{\em Dipolar symmetry and fractonic order.}
%Motivated by above
%realizations, we study particle-defect duality of a 2+1d compact
%Lifshitz model, resembling a gauge dual of a quantum
%smectic. \cite{LRsmecticPRL2020, ZRsmecticAOP2021}
In addition to above examples, Lifshitz model also naturally arises as
the Goldstone-mode (superfluid phase, $\phi$) field theory of the
ordered state of interacting bosons with additional dipole-charge
conservation, explored in great detail in Ref.~\onlinecite{LakeDBHM}.
At harmonic level the symmetry is equivalent to the aforementioned
target-space rotational invariance of a 3d
smectic. \cite{LRsmecticPRL2020, ZRsmecticAOP2021} Our interest in the
Lifshitz model is also motivated by the recent observation that
generalized quantum elastic systems, e.g., 2+1d conventional and
Wigner crystals, supersolids, smectics and vortex crystals, under
elasticity - gauge duality\cite{DasguptaHalperin, FisherLee,
  Zaanen2017} map onto generalized ``fractonic'' gauge
theories\cite{Pretko1,Pretko2}, that exhibit charged matter with
restricted mobility.\cite{PretkoLRdualityPRL2018,
  PretkoLRsymmetryEnrichedPRL2018, PretkoZhaiLRdualityPRB2019,
  KumarPotter19, GromovDualityPRL2019, RHvectorPRL2020,
  LRsmecticPRL2020, ZRsmecticAOP2021}

For concreteness, in what follows, when discussing phases, transitions
and topological defects, I will use the language of bosons,
$\psi\sim e^{i\phi}$ in the dipolar Bose-Hubbard model.\cite{LakeDBHM}
The boson and dipole number conserving symmetry,
\begin{equation}
\phi\rightarrow \phi + \alpha + \bm{\beta}\cdot\xv
\end{equation}
is encoded in the high derivative ``elasticity'' , forbidding lowest
order gradient of the compact superfluid phase $\phi$ (with only
dipole-conserving hopping, e.g.,
$\psi_{\xv+\bm\delta}^\dagger \psi_\xv \psi_\xv
\psi_{\xv-\bm\delta}^\dagger\sim d^\dagger_{\xv,\bm\delta}\psi_\xv
\psi_{\xv-\bm\delta}^\dagger + h.c.$).  The symmetry parameters,
$\alpha, \bm{\beta}$ are zero modes that may be constrained by
system's boundary conditions. The generalized Lifshitz model is a
minimal such continuum field theory, with a Euclidean Lagrangian
density,
\be
\cL = \oh\kappa(\pt_\tau\phi)^2 + \oh K_{ijkl}(\pt_i\pt_j\phi)
(\pt_k\pt_l\phi)^2,
\label{LifshitzL}
\ee
where $\kappa$ is the compressibility, tensor $K_{ijkl}$ encodes
lattice hopping anisotropy and $\tau$ is the $0$-th imaginary time
component of $x_\mu$.  I note that, in striking contrast to the
rotational invariance of the closely-related smectic and other
Lifshitz systems discussed above, here, the more stringent dipolar
symmetry forbids all relevant nonlinearities.\cite{Grinstein86} It
thus protects the fixed line of the noncompact Lifshitz model
\rf{LifshitzL}.

In 2+1d the model \rf{LifshitzL} is generically expected to undergo a
two-stage disordering transition. In the familiar context of smectic
liquid crystals (with $\phi_\xv$ the compact phonon field) it
corresponds to a transition from a smectic state (a periodic array of
stripes, that spontaneously breaks rotational and translational
symmetries, with (quasi-) long-range ordered
($\psi_\xv\sim e^{i\phi_\xv}$)
$d_{\xv,\bm\delta_k}\sim e^{i{\bm\delta_k}\cdot\grad\phi_\xv}\equiv
e^{i\theta_k}$ field), through the translationally-invariant nematic
fluid (that breaks rotational $C_2$ symmetry) to a fully disordered
isotropic and translationally invariant fluid\cite{LRsmecticPRL2020,
  ZRsmecticAOP2021} (with the 3d classical analogue studied for many
decades\cite{ProstDeGennes, ChaikinLubensky}). However, the critical
nature of the nematic-smectic transitions, even in the 3d classical
case\cite{NAtransition} remains an open problem.
%because, until
%recently,\cite{LRsmecticPRL2020, ZRsmecticAOP2021} of the absence of a
%field-theoretic order-parameter formulation.
Here, I utilize duality to provide a gauge theory formulation of the
2+1d Lifshitz model, allowing a transparent characterization of its
phases and a field theoretic analysis of the corresponding Higgs
transitions.

{\em Phases of Lifshitz model.}  To this end, as was introduced in
Refs.~\onlinecite{RHvectorPRL2020, LRsmecticPRL2020,
  ZRsmecticAOP2021}, for a {\em vector} gauge theory formulation of
fractons, it is convenient to reformulate the Lifshitz model in terms
of coupled XY models for the atom ($\psi_\xv\sim e^{i\phi_\xv}$) and
dipole
($d_{\xv,\bm\delta_k}=\psi^\dagger_\xv \psi_{\xv+\bm\delta_k} \sim
e^{i{\bm\delta_k}\cdot\grad\phi_\xv}\equiv e^{i\theta_k}$) superfluid
phases, $\phi$ and $(\thv)_k = \theta_k$, with a Lagrangian
density,\cite{compactComment}
\bea
\cL &=& 
\oh\kappa(\pt_\tau\phi)^2
+\oh g(\grad\phi + \thv)^2
+\oh I(\pt_\tau\thv)^2\nonumber\\
&&+ \oh K_{ijkl}(\pt_i\theta_j)(\pt_k\theta_l)\ .
\label{LifshitzXYcoupled}
\eea
At low energies the $g$ coupling in $\cL$ enforces
$\grad\phi \approx -\thv$ (i.e.,
$\phi \approx\phi_0 -{\bm\thv}\cdot\rv$) and thus reduces $\cL$
\rf{LifshitzXYcoupled} to the standard form in \rf{LifshitzL}, with
corrections that are subdominant at low energies.  This form of
Lifshitz Lagrangian \rf{LifshitzXYcoupled} displays a gauge-like
coupling between atoms and dipoles, that thereby underlies a
nontrivially intertwined atom-dipole (and corresponding vortices)
dynamics of the Lifshitz fluid and its aforementioned phase
transitions.\cite{commentLake}

Before turning to a detailed analysis, \rf{LifshitzXYcoupled} already
reveals the structure of the phases of the Lifshitz model:

\noindent (i) In the absence of vortices, i.e., a fully Bose-condensed
state of atoms and dipoles, BEC$_{ad}$ is characterized by
single-valued $\phi$ and $\thv$ phases. The state is gapless and is
well-described by a Gaussian fixed line of standard Lifshitz form
\rf{LifshitzL}, with a dynamical exponent $z=2$.  For constant $\thv$,
the BEC$_{ad}$ state is orientationally ordered, atomic condensate at
momentum $\thv$ akin to a Fulde-Ferrell\cite{FF, LO,
  LR_VishwanathPRL,LRpra}, a spin-orbit coupled
condensate\cite{LR_ChoiPRL,HuiZhai} and a helical state of frustrated
bosons\cite{helicalBosonsSMR22}.  However, I expect it to be
challenging to probe this momentum, since in the bulk it can be gauged
away.\cite{KRnoncentrosymmSC} Given the resemblance of
\rf{LifshitzXYcoupled} to the Abelian-Higgs model (with a non-gauge
invariant ``Maxwell'' sector for $\thv$, characterized by $K_{ijkl}$),
I expect the BEC$_{ad}$ - BEC$_{d}$ transition to be in a generalized
normal-superconductor universality class.  This is expected due to a
nontrivial gauge-like coupling between the dipolar and atomic
condensates, whose consequences we will also see in the dual gauge
theory formulation descussed below.

\noindent (ii) Increasing fluctuations (at zero temperature done by
increasing boson interaction relative to dipole hopping), drives a
proliferation of vortices in the atomic phase $\phi$ Mott-insulating
atoms, with dipoles remaining Bose-condensed in BEC$_{d}$, and in the
case of an underlying isotropic system spontaneously breaks rotational
symmetry by the choice of $\thv$.  With this $\grad\phi$ becomes an
independent vector field (with both transverse and longitudinal
components) that can thus be safely integrated out, leading to a $z=1$
XY-like Lagrangian density for the dipolar Goldstone mode $\th$,
\be
\cL_{BEC_d}  = \oh I(\pt_\tau\thv)^2
+ \oh K_{ijkl}(\pt_i\theta_j)(\pt_k\theta_l)\ .
\label{LifshitzXYcoupledBECd}
\ee

\noindent (iii) Increasing interaction further then proliferates vortices in
$\thv$, leading to a fully Mott-insulating phase, MI of atoms and
dipoles.

\begin{figure}[htbp]
  \hspace{-0.13in}\includegraphics*[width=0.5\textwidth]{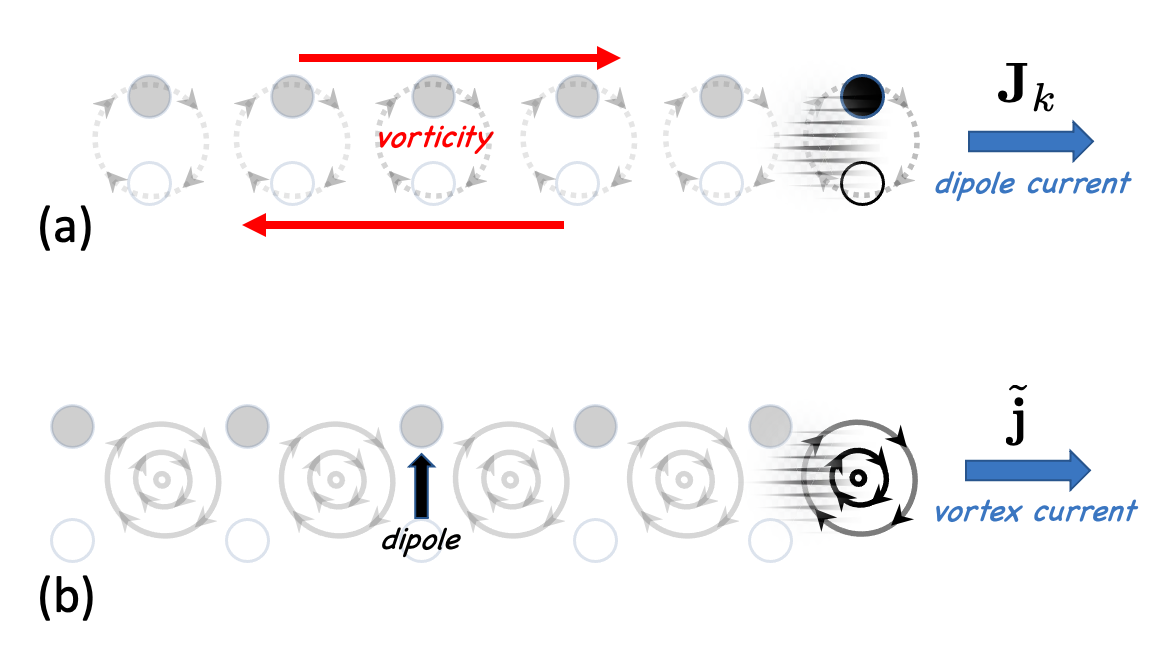}
  \caption{(a) An illustration of a dipole current ${\bf J}_k$
    transverse to the dipole moment and its equivalence to atomic
    counter-flow (red arrows) and the associated induced vortex
    density $\tilde j_0$. They both contribute to atomic circulation,
    corresponding to dual charge density of the Gauss's law
    \rf{dualLifshitzGauss}. (b) An illustration of a vortex current
    ${\bf j}$ and induced transverse dipole density (black arrow),
    $J_{0k}$, both contributing to atomic number imbalance, as dual
    current sources of the Ampere's law \rf{dualLifshitzAmpere}.}
\label{DipoleVortexCurrentFig}
\end{figure}

The shortcoming of the continuum form \rf{LifshitzXYcoupled} of the
compact Lifshitz model, $\cL$, is that compactness (i.e., vortex
degrees of freedom) of the Goldstone modes $\phi$ and $\thv$ is not
manifest.\cite{VillainComment} To remedy this, I make the
corresponding vortex degrees of freedom explicit by allowing
nonsingle-valued configurations of $\phi$ and $\thv$.  Namely, I
``gauge'' $\cL$ in \rf{LifshitzXYcoupled}, with atomic (a-) and
dipolar (d-) vortices, respectively represented by fluxes of the
associated gauge fields,\cite{gaugingComment}
\bea
\cL &=& \cL_{\text M}[\tilde a_\mu,\tilde\A_\mu] + \oh\kappa(\pt_\tau\phi - \tilde a_0)^2
+\oh g(\grad\phi - \tilde\av + \thv)^2\nonumber\\
&+&\oh I(\pt_\tau\thv - \tilde\A_0)^2
+ \oh K_{ijkl}(\pt_i\theta_j - \tilde A_{ij})(\pt_k\theta_l - \tilde
A_{kl}),\nonumber\\
&=& \cL_{\text M}[\tilde a_\mu,\tilde\A_\mu]
+ \frac{\kappa}{2}(\pt_\mu\phi - \tilde a_\mu + \theta_\mu)^2
+\frac{K}{2}(\pt_\mu\thv - \tilde\A_\mu)^2.\nonumber\\
\label{LifshitzGaugetheory}
\eea
with the corresponding {\em discrete} vortex (dual) 3-currents given
by
\bea
\tilde j_\mu &=& \epsilon_{\mu\nu\gamma}\pt_\nu\tilde a_\gamma
- \epsilon_{\mu\nu k}\tilde A_{\nu k}
\equiv(\pt \times \tilde a)_\mu - \tilde A^*_{\mu} ,\nonumber\\
&=& \sum_p\int_{\tau_p} n_p
\hat{v}_\mu(\tau_p)\delta^3(x^\nu - x^\nu_p(\tau_p)),
\label{jtilde}\\
\tilde \J_\mu &=& \epsilon_{\mu\nu\gamma}\pt_\nu\tilde \A_\gamma 
\equiv(\pt \times \tilde \A)_\mu,\nonumber\\
&=& \sum_p\int_{\tau_p} {\bf N}_p
\hat{V}_\mu(\tau_p)\delta^3(x^\nu - x^\nu_p(\tau_p)),
\label{Jtilde}
\eea
$\theta_\mu = (0,\theta_i)$,
$\tilde a_\mu = (\tilde a_0, \tilde a_i)$,
$(\tilde \A_\mu)_k =\tilde A_{\mu k} = (\tilde A_{0 k}, \tilde A_{i
  k})$, $(n_p, {\bf N}_p)$ (boson, dipole) $p$-th vortex integer
windings, and $(\hat{v}_\mu(\tau_p), \hat{V}_\mu(\tau_p))$ unit
3-velocities of their corresponding world-lines.  Throughout, to
emphasize the structure of the expressions I use a short-hand
notation:(i) bold-faced for Roman flavor $k$ and spatial $i, j$
indices, and, (ii) where obvious, omit the space-time Greek indices,
as defined below.  In the last line in \rf{LifshitzGaugetheory}, for
transparency of analysis I took $K_{ijkl} = K\delta_{ik}\delta_{jl}$
and rescaled coordinates so that $g = \kappa$ and $I = K$, i.e., chose
the speeds of ``sound'' to be $1$; in an isotropic lattice-free system
$K_{ijkl}$ reduces to a tensor of Frank elastic
energy\cite{ProstDeGennes} for $\thv$, characterized by twist, splay
and bend elastic constants, that, more generally will be broken by the
underlying lattice.  With a- and d-vortices encoded by gauge fields
$a_\mu, A_{\mu k}$, above phases are single-valued Goldstone modes
satisfying,
$\epsilon_{\mu\nu\gamma}\pt_\nu\pt_\gamma\phi\equiv\pt \times\pt\phi =
0$,
$\epsilon_{\mu\nu\gamma}\pt_\nu\pt_\gamma\thv\equiv\pt \times\pt\thv =
0$, where I have used the same symbols for simplicity of presentation.
In \rf{LifshitzGaugetheory} I also included vortex 3-current core
energies (accounting for the lattice-scale physics), that take the
form of a generalized Maxwell Lagrangian,
\bea
\cL_{\text M} &=&\oh E_{c1}(\pt\times \tilde A_k)^2
+\oh E_{c2}((\pt\times \tilde a)_\mu
- \tilde A^*_\mu)^2,
\label{LifshitzMaxwell}
\nonumber\\
\eea
where I defined a Hodge-dual of $\tilde A_{\nu k}$ as
$\tilde A^*_{\mu} \equiv \epsilon_{\mu\nu k}\tilde A_{\nu k}$.
$\cL$ is invariant under a generalized gauge transformation,
\bea
\phi\rightarrow\phi + \alpha,&&\quad
\thv\rightarrow\thv + \chiv, \label{genGauge}\\
\tilde a_\mu \rightarrow \tilde a_\mu + \pt_\mu\alpha
+ \delta_{\mu j}\chi_j,&&\quad
\tilde\A_\mu \rightarrow \tilde\A_\mu + \pt_\mu\chiv.\nonumber
\eea
and in \rf{LifshitzMaxwell} the flux $(\pt\times\tilde a)_\mu$ is
itself gauged by a 2-form component of $\tilde\A_\mu$, according to
$d\tilde a - \tilde A$.  This reflects arbitrariness of division
between dipole current and atom vorticity, as illustrated in
Fig.~\ref{DipoleVortexCurrentFig}a.  Microscopically this corresponds
to the contribution of the antisymmetric component of the dipole
current $\grad\times\thv = \epsilon_{ik}\tilde A_{ik}$ to the bosonic
vortex $\grad\times\tilde\av$, encoded in the combination
$\grad\phi + \thv$.

With discrete atom, $n_p$ and dipole, ${\bf N}_p$ vortex charges,
\rf{jtilde}, \rf{Jtilde} the Lifshitz model (as the aforementioned 3d
classical smectic\cite{ProstDeGennes, ChaikinLubensky,
  LRsmecticPRL2020, ZRsmecticAOP2021}) displays three phases of a
dipole-conserving bosonic fluid\cite{LakeDBHM}:

\noindent (1) MI: a gapped phase with proliferated {\em both} atomic
and dipole vortices, thereby described by {\em continuous}
$\tilde a_\mu\; \tilde\A_\mu$ gauge fields, with a gapped Debye-Huckel
Lagrangian density
\bea
\cL_{\text{MI}}
%&=& \cL_{\text M}[\tilde a_\mu,\tilde\A_\mu]
%+ \oh\kappa\tilde a_0^2
%+\oh g(\tilde\av - \thv)^2\nonumber\\
%&+&\oh \tilde I \tilde\A_0^2
%+ \oh K_{ijkl}\tilde A_{ij}\tilde A_{kl},\\
%&=& \oh \tilde I \tilde\A_0^2 
%+ \oh \tilde K_{ijkl}\tilde A_{ij}\tilde A_{kl}
%+\oh E_{c1}(\pt\times \tilde A_k)^2,\qquad
&=& \frac{K}{2}(\tilde{\bf A}_\mu)^2 
+ \frac{E_{c1}}{2}(\pt\times \tilde A_k)^2\nonumber\\
&& +\frac{\kappa}{2}\tilde a_\mu^2 + \frac{E_{c2}}{2}((\pt\times \tilde a)_\mu
- \tilde A^*_\mu)^2,
\label{LifshitzGaugetheoryMI}
\eea
where $\tilde {\bf A}_\mu$ and $\tilde a_\mu$ are implicitly
understood as gauge invariant
%\equiv \epsilon_{\mu\nu\gamma} \pt_\nu
%A_{\gamma k}$
projections transverse to momentum
$k_\nu$, and in this vortex condensate phase I used \rf{genGauge} to
gauge away the phases,
$\phi,\thv$, with a transverse low-energy constraint $\theta \approx
\tilde a$.

\noindent (2) BEC$_d$: a gapless,
$z=1$, orientationally ordered condensate of dipoles (vacuum of dipole
vortices $\tilde\A_\mu =
0$) and MI of atoms. This gapless insulator is characterized by a
Lagrangian density
\bea
\cL_{\text{BECd}} &=&
%\oh I (\pt_\tau\tilde\av)^2 
%+ \oh K_{ijkl}(\pt_i\tilde a_j)(\pt_k\tilde a_l) 
%+ \oh E_{c2}(\pt\times\tilde a)^2.\nonumber\\
\oh K (\pt_\mu{\thv})^2 + \oh E_{c2}(\pt\times \theta)^2,
\label{LifshitzGaugetheoryBECd}
\eea
with a low-energy transverse constraint $\theta \approx\tilde a$.
% $\pt\times\theta = \pt\times\tilde
% a$.  $\epsilon_{\mu\nu\gamma}\partial_\nu\tilde
% a_\gamma\approx\epsilon_{\mu\nu\gamma}\partial_\nu\theta_\gamma$.

\noindent (3) BEC$_{ad}$: a gapless, $z=1$, orientationally ordered
condensate of atoms and dipoles, characterized by vanishing gauge
fields $\tilde a_\mu = \tilde\A_\mu = 0$, with a vortex-free
Lagrangian density, $\cL_{\text{BECad}}$ at low energies is given by
\rf{LifshitzXYcoupled} and equivalently by \rf{LifshitzL}.

Although above description of the phases MI, BEC$_d$, BEC$_{ad}$ is
quite transparent in this picture, because in \rf{jtilde},\rf{Jtilde}
the gauge fields are sourced by {\em discrete} vortex charges, the
nature of the MI-BEC$_d$ and BEC$_d$-BEC$_{ad}$ quantum phase
transitions (beyond a mean-field approximation) is not easily
accessible. In contrast, a dual gauge theory provides a suitable field
theory of these transitions.

{\em Lifshitz boson-vortex duality.}  Following quantum smectic
studies\cite{RHvectorPRL2020,LRsmecticPRL2020,ZRsmecticAOP2021} I
dualize the Lifshitz model, obtaining a gauge theory that encodes the
restricted mobility of its charge and dipole vortices and a provide a
description of its phase transitions. To this end I introduce a
Hubbard-Stratonovich atom and dipole currents, $j_\mu, J_{\mu k}$ and
integrate out the smooth component of the superfluid phases
$\phi, \theta_k$, that impose atom and dipole conservation constraint,
\be
\pt_\mu j_\mu = 0,\quad \pt_\mu J_{\mu k} = j_k,
\label{continuity}
\ee
latter encoding that motion of atoms generates dipoles. These are
respectively solved by gauge fields, $a_\mu, A_{\mu k}$ with
%
%\be
%j_\mu = \epsilon_{\mu\nu\gamma}\pt_\nu a_\gamma,\quad
%J_{\mu k} = \epsilon_{\mu\nu\gamma}\pt_\nu A_{\gamma k}
%+ \epsilon_{\mu\nu k} a_\nu,
%\ee
%
%
\bea
j_\mu &=& \epsilon_{\mu\nu\gamma}\pt_\nu a_\gamma
\equiv (\pt\times a)_\mu,\\
J_{\mu k} &=& \epsilon_{\mu\nu\gamma}\pt_\nu A_{\gamma k}
+ \epsilon_{\mu\nu k} a_\nu\equiv(\pt\times A_k + a^*_k)_\mu,
\label{jJsolns}
\eea
and allows the interpretation of \rf{continuity} as generalized
coupled Faraday equations,
\be
\grad\times{\bf e} = -\pt_\tau b,
\quad
\grad\times{\bf E}_k = -\pt_\tau B_k + \epsilon_{k i} e_i\, ,  
\label{Faraday}
\ee
for the electric and magnetic fields,
$e_i = -\epsilon_{ij} j_j, b = j_0, E_{ik} = -\epsilon_{ij} J_{jk} , B_k
= J_{0k}$,
\bea
\hspace*{-2cm}
b &=& \grad\times \av,\;\;\ev = -\pt_\tau \av + \grad a_0,\\
B_k &=& \grad\times \A_k  - \hat\xv_k\times {\bf a},\;\;\E_{k}=-\pt_\tau
\A_{k} + \grad A_{0 k} - \hat\xv_k a_0.\qquad\hspace*{-1cm}\nonumber
\eea
$\hat\xv_k$ is unit coordinate vector with components $\delta_{ik}$. I
note that $\epsilon_{k i} e_i$ (atomic current) appears as the
magnetic monopole current sourcing the dipole Faraday equation
\rf{Faraday}. Above dual field strengths and currents are invariant
under generalized dual gauge transformation
\bea
%\phi\rightarrow\phi + \alpha,&&\quad
%\thv\rightarrow\thv + \chiv,\\
 a_\mu \rightarrow  a_\mu + \pt_\mu\tilde\phi\;,
&&
 A_{\mu k} \rightarrow A_{\mu k} + \pt_\mu\theta_k - \delta_{\mu k}\tilde\phi\;.\;\;
\label{dualGenGauge}
\eea
With this, the Lifshitz model \rf{LifshitzL}, \rf{LifshitzXYcoupled},
\rf{LifshitzGaugetheory} transforms to a generalized mutual
Chern-Simons-Maxwell Lagrangian,
\be
\cL  =
\cL_{\text M}[\tilde A_{\mu k},\tilde a_\mu] + \cL_{\text{CS}}[\tilde A_{\mu k},\tilde
a_\mu, A_{\mu k}, a_\mu] + \tilde \cL_{\text
  M}[A_{\mu k}, a_\mu],
\label{almostDualL}
\ee
where
$\cL_{\text{CS}} = i \tilde A_{\mu k} J_{\mu k}+ i \tilde a_\mu j_\mu
= i A_{\mu k} \tilde J_{\mu k} + i a_\mu \tilde j_\mu$ is the coupling
of atoms and dipoles to the associated a- and d-vortices,
\bea
\cL_{\text{CS}} &=&
  i\tilde A_{k}\cdot(\pt \times A_k + a_k^*)
  + i \tilde a\cdot \pt\times a,\\
  &=& i a\cdot(\pt\times\tilde a - \tilde A^*_{k})
  + i A_k\cdot \pt \times\tilde A_k,
\label{dualCSLifshitz}
\eea
\\
with $(a_{k}^*)_\mu\equiv \epsilon_{\mu\nu k} a_\nu$, $\tilde A^*_{\mu}
\equiv \epsilon_{\mu \nu k}\tilde A_{\nu k}$, type of Hodge-duals of
$a_\nu$ and $\tilde A_{\mu k}$, and 
\bea
\tilde\cL_{\text M} &=&
%\oh\tilde K^{-1}_{ijkl}(E_{ij} - \delta_{ij}a_0)
%(E_{kl} - \delta_{kl}a_0) + \oh I^{-1}(B_k - \epsilon_{ki} a_i)^2
%\nonumber\\
%&&+\oh\kappa^{-1} b^2 +\oh g^{-1} \ev^2.
\oh K^{-1}(\pt\times  A_k + a_k^*)^2
+\oh \kappa^{-1}(\pt\times a)^2,
\label{LifshitzDualMaxwell}
\eea
the generalized Maxwell Lagrangian for bosons dual to $\cL_{\text{M}}$
\rf{LifshitzMaxwell}.  I note the appealing symmetric form between
bosonic matter and corresponding vortices.

To complete duality, I express $\cL$ above in terms of vortex
currents,
\bea
\tilde\cL &=&
\oh K^{-1}(\pt\times  A_k + a_k^*)^2
+\oh\kappa^{-1}(\pt\times a)^2\nonumber\\
&&+ i {\bf A}_{\mu}\cdot \tilde{\bf  J}_{\mu} + i a_\mu \tilde j_\mu
+ \frac{E_{c1}}{2} \tilde{\bf J}_\mu^2 + \frac{E_{c2}}{2} \tilde
j_\mu^2\;,
\label{dualLajAJ}
\eea
and sum over these discrete vortex currents, obtaining dual
Lagrangian, $\tilde\cL$,
\bea \tilde\cL&=& \oh K^{-1}(\pt\times A_k + a_k^*)^2
+\oh\kappa^{-1}(\pt\times a)^2 \label{dualLmaster}\\
&-&\tilde K_a\cos(\pt_\mu\tilde\phi - a_\mu) -\tilde
K_d\cos(\pt_\mu\tilde\theta_k - \delta_{\mu k}\tilde\phi -
A_{\mu k}),\nonumber
\eea
where I utilized \rf{dualGenGauge} to include dual matter (vortex)
degrees of freedom, $\tilde\phi,\tilde\thv$, and for transparency of
presentation approximated the Villain potential by its lowest
harmonic. The compact dual phases are subject to integer winding
boundary conditions along $\tau$,
$\tilde\phi_\xv(\tau+\beta) = \tilde\phi_\xv(\tau) + 2\pi w_\xv$,
$\tilde\thv_\xv(\tau+\beta) = \tilde\thv_\xv(\tau) + 2\pi {\bf
  w}_\xv$, $w_\xv\in\mathbb{Z}$, ${\bf w}_\xv\in\mathbb{Z}$.

As required, the dual Lifshitz model reproduces the three phases
discussed above:

\noindent (1) MI: a gapped condensate of dual (atom and dipole vortex)
matter, that Higgses gauge fields $a_\mu, A_{\mu k} $ that encode
bosonic and dipole matter, thereby fully gapping them. The resulting
Lagrangian density is
\bea
\tilde\cL_{\text{MI}} &=& \oh K^{-1}(\pt\times A_k + a_k^*)^2
+\oh\kappa^{-1}(\pt\times a)^2\nonumber\\
&-&\tilde K_a\cos(a_\mu) -\tilde
K_d\cos(A_{\mu k}),
\label{dualLifshitzGaugetheoryMI}
\eea
a dual of \rf{LifshitzGaugetheoryMI}.

\noindent (2) BEC$_d$: an orientationally-ordered, gapless, $z=1$
state, that is a dual condensate of atomic vortex matter, Higgsing
$a_\mu$ and an insulator of dipole vortex matter (that thereby
decouples), giving a Lagrangian density,
\bea
\tilde\cL_{\text{BEC}_{d}} &=& \oh K^{-1}(\pt\times A_k + a_k^*)^2 
+\oh\kappa^{-1}(\pt\times a)^2\nonumber\\
&-&\tilde K_a\cos(a_\mu),\\
 &\approx& \oh K^{-1}(\pt\times A_k)^2,
\label{dualLifshitzGaugetheoryBECd}
\eea
a dual of \rf{LifshitzGaugetheoryBECd}.

\noindent (3) BEC$_{ad}$: a gapless, $z=2$ state that is a dual
insulator of a- and d-vortex matter, $\tilde\phi,\tilde\theta_k$,
(allowing one to set $\tilde J_{\mu k} =\tilde j_\mu =0$ in
\rf{dualLajAJ}), leading to a dual Maxwell Lagrangian,
$\tilde\cL_{\text{BEC}_{ad}} = \tilde\cL_{\text M}$,
\rf{LifshitzDualMaxwell}. It is a dual to the Lifshitz superfluid
state, with a Lagrangian \rf{LifshitzXYcoupled} and \rf{LifshitzL}.

In this dual picture the BEC$_{ad}$-to-BEC$_{d}$ transition is driven
by a condensation of atomic vortices,
$\tilde\psi\sim e^{i\tilde\phi}$, with an insulating (vacuum) state of
dipole vortex matter, $\tilde{d}_k\sim e^{i\tilde{\theta}_k}$
(corresponding to a dipole condensate). The latter property decouples
disordered d-vortex matter (last term in \rf{dualLmaster}), allowing
one to integrate it out. With this observation, the
BEC$_{ad}$-to-BEC$_{d}$ transition is thus described by a generalized
Abelian-Higgs model (a dual superconductor), with the Lagrangian
density,
\bea \tilde\cL&=& \oh K^{-1}(\pt\times A_k + a_k^*)^2
+\oh\kappa^{-1}(\pt\times a)^2\nonumber\\
&+&\frac{\tilde
  K_a}{2}|(\pt_\mu - i a_\mu)\tilde\psi|^2 + V_a(|\tilde\psi|),
 \label{dualLglBECad}
\eea
where $V_a(|\tilde\psi|)$ is a standard $U(1)$-symmetric Landau
potential for atomic vortex matter.  In the BEC$_d$ the dual a-vortex
condensate $\tilde\psi$ thus Higgses out $a_\mu$ (quantizing
Mott-insulating atomic matter) giving a gapless Maxwell dipole
Lagrangian for $A_{\mu k}$, \rf{dualLifshitzGaugetheoryBECd}.

The subsequent BEC$_{d}$-to-MI transition is then driven by a
condensation of dipole vortices, $\tilde d_k\sim e^{i\tilde\theta_k}$
from a condensed (Higgs) BEC$_{d}$ state of atomic vortex matter (with
a gapped atomic gauge field $a_\mu$).  With this, the BEC$_{d}$-to-MI
transition is thus described by a conventional Abelian-Higgs model,
with the Lagrangian density,
\bea \tilde\cL_{\text{BEC$_{d}$-MI}}&=& \oh K^{-1}(\pt\times A_k)^2 +\frac{\tilde
  K_d}{2}|(\pt_\mu - i A_{\mu k})\tilde d_k|^2 \nonumber\\
&+& V_d(|\tilde d_k|),
\label{dualLglBECd}
\eea
where $V_d(|\tilde d_k|)$ is a standard $U(1)$-symmetric Landau
potential for dipole vortex matter.  In the MI the dual d-vortex
condensate $\tilde d_k$ thus Higgses out $A_{\mu k}$ (quantizing
Mott-insulating dipolar matter) giving a fully gapped Lagrangian for
$a_\mu$ and $A_{\mu k}$, \rf{dualLifshitzGaugetheoryMI}.  I note that
generically the two flavors of the $k=x, y$ dipoles may condense at
two distinct transitions, allowing for yet another intermediate phase,
where only one of the $d_x$ and $d_y$ has
condensed.\cite{LRsmecticPRL2020,ZRsmecticAOP2021, LakeDBHM}

One appeal of above dual description is that the BEC$_{ad}$-BEC$_d$
and BEC$_d$-MI quantum phase transitions are Higgs transitions
(associated with condensation of atomic vortex and dipole vortex
matter, respectively), well described by conventional gauge field
theories \rf{dualLglBECad}, \rf{dualLglBECd}.  Thus duality allows a
computation of criticality beyond a mean-field approximation. I leave
these nontrivial analyses for future studies.

The corresponding dual Hamiltonian is given by,
\bea
\tilde\cH_{\text M} &=&
\frac{K}{2}\E_k^2 + \frac{K}{2}(\grad\times\A_k  - \hat\xv_k\times {\bf a})^2
+\frac{\kappa}{2}\ev^2 + \frac{\kappa}{2} (\grad\times\av)^2\nonumber\\
&&+ i \A_k\cdot\tilde\J_k +i\av \cdot\tilde\jv,
\label{LifshitzDualMaxwellH}
\eea
with canonically conjugate electric fields and gauge potentials. The
associated coupled Gauss's laws,
\bea
\grad\cdot\ev = \tilde j_0 - E_{ii}\;,\quad
\grad\cdot\E_k = \tilde J_{0k},
\label{dualLifshitzGauss}
\eea
encode a relation between circulations of atomic and dipolar currents
and the corresponding vortex densities, $\tilde j_0, \tilde J_{0k}$.
The appearance of $E_{ii}$ as a source of the atomic Gauss's law
correctly encodes the dipolar current
$\epsilon_{ik}J_{ik} \sim {\bf d} \times {\bf v}$ transverse to the
local dipole moment $ {\bf d}$, a bosonic counter-flow that
contributes to atomic vorticity, as illustrated in
Fig.\ref{DipoleVortexCurrentFig}(a).

Finally, to further elucidate Lifshitz model dynamics, I examine the
atomic and dipole Ampere's equations (in real time),
\bea
\kappa^{-1}\grad\times b &=& -\pt_t {\bf e} + \tilde {\bf j}
-\kappa^{-1} {\bf B}^*,\label{dualLifshitzAmpere}\\
K^{-1}\grad\times B_k &=& -\pt_t {\bf E}_k + \tilde {\bf J}_k,\label{AmpereA}
\eea
corresponding to Lagrangian \rf{dualLajAJ}.  In
\rf{dualLifshitzAmpere} I note that the vortex current
$\tilde {\bf j}$ induces a dipole density
${\bf B}^*_i \equiv \epsilon_{i k} B_k$ and a gradient in atom density
$\grad\times b$ transverse to the vortex current. The detailed
physical content of this intriguing relation is illustrated in
Fig.\ref{DipoleVortexCurrentFig}(b).  I also note that in terms of
atomic phase $\phi$, the source-free atomic Ampere's law just
corresponds to a vortex-free condition
$\grad\pt_t\phi = \pt_t\grad\phi$.  This condition is violated by a
vortex current $\tilde {\bf j}$ and dipole density ${\bf
  B}^*_i$. Equivalently, in terms of atom and dipole densities, in
steady state Ampere's law simply corresponds to force balance between
a gradient of the atomic chemical potential (to lowest order the
atomic density $n$), a dipole density and the Magnus force associated
with the vortex current. I thus observe that Gauss's and Ampere's
laws illustrate boson-dipole cross-coupling and the associated vortex
defects, encoded in the Lagrangian \rf{LifshitzDualMaxwellH},
\rf{dualLifshitzGauss}.

%\begin{widetext}
%\begin{figure}[htbp]
%  \hspace{-0.13in}\includegraphics*[width=0.5\textwidth]{phasediagram}
%  \caption{Illustration of quantum melting of a 2D crystal into a
%    smectic, followed by smectic-to-nematic melting, respectively
%    driven by a condensation of $b_x$ and of $b_y$ dislocations.}
%  \label{fig:phase transition}
%\end{figure}
%\end{widetext}

{\em Summary:} In this manuscript I studied phases and phase
transitions of a quantum $2+1$d Lifshitz model, a continuum
Goldstone-mode field theory of a Bose Hubbard model with dipole
conservation.\cite{LakeDBHM} Reformulating the dipole-conserving
second derivative coupling in terms of coupled XY-models of bosons and
their dipoles, allows for a description of the phases and transitions
in terms of an extension of familiar Bose-condensed and Mott-insulator
phases of bosons and dipoles.  I complement this direct analysis by a
dual coupled gauge theory, that elucidates nontrivial dynamics
between bosons, dipoles and their corresponding vortices. It also
allows for a transparent description of these transitions as
generalized Higgs transitions, whose beyond-mean-field criticality I
leave for future studies.

{\em Note Added:} After this work was completed I received an
interesting preprint from Pranay Gorantla, {\em et al.}, presenting a
complementary lattice duality of a 2+1d compact Lifshitz model, with a
detailed treatment of the ground state degeneracy for the periodic
boundary conditions.\cite{ShuHengLifshitz}

\emph{Acknowledgments}.  I thank Shu-Heng Shao and Nati Seiberg for
illuminating discussions and for sharing their manuscript before its
submission. I appreciate Anton Kapustin's comments on the manuscript
and acknowledge support by the Simons Investigator Award from The
James Simons Foundation.


\begin{thebibliography}{99}
\bibitem{mLifshitz} A generalized $m$-Lifshitz model, has elasticity
  with $m$ ``soft'' (Laplacian) axes and $d-m$ complementary ``hard''
  (gradient) axes. In this nomenclature, the conventional classical 3d
  smectic is the $m=2$-Lifshitz model, and 2+1d quantum smectic and 3d
  classical columnar liquid crystal are described by the
  $m=1$-Lifshitz model.  Other generalizations include a nonscalar
  Goldstone mode field as in e.g., tethered
  membranes\cite{JWSmembranes,LRmembranes, RTtubulePRL, RTtubulePRE}
  and nematic elastomers\cite{XRelastomers}
\bibitem{compactComment} It is clear from the definition of the boson
  $\psi_\xv\sim e^{i\phi_\xv}$) and dipole
  $d_{\xv,\bm\delta_k}=\psi^\dagger_\xv \psi_{\xv+\bm\delta_k} \sim
  e^{i{\bm\delta_k}\cdot\grad\phi_\xv}\equiv e^{i\theta_k}$ operators
  that $\phi_\xv$ and
  $\theta_{k,\xv} = \phi_{\xv+\bm\delta_k} - \phi_{\xv}\approx
  {\bm\delta}_k\cdot\grad\phi_\xv$ are compact angular fields. On a
  square lattice, where the two lattice vectors ${\bm\delta}_k$ have
  components $a\delta_{ik}$ the minimal coupling reduces to
  $\oh g (a\grad\phi - \thv)^2$ (as usual understood to be regularized
  by lattice differences inside periodic potential e.g.,
  $- g\cos(a\grad\phi - \thv)$\cite{VillainComment}, corresponding to
  the effective charge $1/a$, that for simplicity of presentation I
  took to be $1$.
\bibitem{NAtransition} The 3d classical nematic-to-smectic phase
  transition is one of the most enigmatic ones, remaining an open
  problem, despite extensive
  theoretical\cite{deGennes72,HLM,NelsonToner, Lubensky81,TonerNSm,
    LRsmecticAerogel} and experimental\cite{GarlandExps} studies.
\bibitem{VillainComment} With care of treating compactness of $\phi$
  and discreteness of its defects, the continuum approach gives
  equivalent results to a lattice formulation with compactness encoded
  via a Villain potential.
\bibitem{gaugingComment} I thus note that ``gauging'' a model with
  compact global symmetry thereby takes on a physical content, by
  making its compactness manifest.  The associated field strength is
  vortex charge and current densities, and gauge redundancy is simply
  the arbitrariness of division between longitudinal (Goldstone-mode)
  and transverse (vortex) degrees of freedom.
\end{thebibliography}
\end{document}